\documentclass[conference]{IEEEtran}
\IEEEoverridecommandlockouts
\usepackage{cite}
\usepackage{amsmath,amssymb,amsfonts}
\usepackage{algorithmic}
\usepackage{graphicx}
\usepackage{textcomp}
\usepackage{xcolor}
\def\BibTeX{{\rm B\kern-.05em{\sc i\kern-.025em b}\kern-.08em
    T\kern-.1667em\lower.7ex\hbox{E}\ki have a latex dern-.125emX}}

\usepackage[normalem]{ulem}      % for strikeout (\sout{}) without affecting \emph

\usepackage{pdfpages}

\begin{document}

\title{LLM-Powered Virtual Patient Agents for Interactive Clinical Skills Training with Automated Feedback}
%*\\
%{\footnotesize \textsuperscript{*}Note: Sub-titles are not %captured in Xplore and

%\author{Anonymous Submission}

\author{
\IEEEauthorblockN{Henrik Voigt}
\IEEEauthorblockA{\textit{Department of Computer Science} \\
\textit{University of Jena}\\
Jena, Germany \\
henrik.voigt@uni-jena.de}

\and

\IEEEauthorblockN{Yurina Sugamiya}
\IEEEauthorblockA{\textit{Department of Robotics} \\
\textit{Tokyo Denki University}\\
Tokyo, Japan \\
sugamiya@aoni.waseda.jp}

\and

\IEEEauthorblockN{Kai Lawonn}
\IEEEauthorblockA{\textit{Department of Computer Science} \\
\textit{University of Jena}\\
Jena, Germany \\
kai.lawonn@uni-jena.de}

\and

\IEEEauthorblockN{Sina Zarrieß}
\IEEEauthorblockA{\textit{Department of Computational Linguistics} \\
\textit{Bielefeld University}\\
Bielefeld, Germany \\
sina.zarriess@uni-bielefeld.de}

\and

\IEEEauthorblockN{Atsuo Takanishi}
\IEEEauthorblockA{\textit{Department of Robotics} \\
\textit{Waseda University}\\
Tokyo, Japan \\
takanisi@waseda.jp}

}

\maketitle

\begin{abstract}
Objective Structured Clinical Examinations (OSCEs) are essential for medical training, but they require significant resources, including professional actors and expert medical feedback. Although Large Language Models (LLMs) have introduced text-based virtual patients for communication practice, these simulations often lack the capability for richer, non-textual interactions. This paper presents a novel framework that significantly enhances LLM-based simulated patients by equipping them with action spaces, thereby enabling more realistic and dynamic patient behaviors that extend beyond text. Furthermore, our system incorporates virtual tutors that provide students with instant, personalized feedback on their performance at any time during these simulated encounters. We have conducted a rigorous evaluation of the framework's real-time performance, including system latency and component accuracy. Preliminary evaluations with medical experts assessed the naturalness and coherence of the simulated patients, as well as the usefulness and appropriateness of the virtual tutor's assessments. This innovative system provides medical students with a low-cost, accessible platform for personalized OSCE preparation at home.
\end{abstract}


%%%%%%%%%%%%%%%% SECTIONS %%%%%%%%%%%%%%%%%%%%%%%%%%%%%%%%%

\section{INTRODUCTION}
\label{sec:introduction}
Objective Structured Clinical Examinations (OSCEs) play a key role in medical education, providing a means of assessing students' clinical competencies in a variety of scenarios. However, ensuring that students receive sufficient and effective OSCE practice poses a significant challenge. Traditional training requires substantial resources, such as professional actors to simulate patients and medical experts to provide real-time feedback, which makes frequent, personalized practice costly and difficult to implement on a large scale.

Various simulation technologies have been explored in medical education to mitigate these challenges. These range from digital patient simulators to physical robotic patients, which offer more realistic and engaging training by physically embodying patients. Using simulated patients can reduce reliance on human actors and associated costs. However, two major limitations persist. Firstly, many existing patient simulators, particularly those leveraging recent technological advances, are limited to text-based interactions. Secondly, simulating the nuanced role of an OSCE examiner — especially generating high-quality, comprehensive feedback that goes beyond simple keyword matching for complex clinical skills — remains a significant hurdle. Current automatic scoring systems often rely on predefined state graphs, which struggle to evaluate intricate skills such as thorough symptom gathering.

The recent emergence of powerful LLMs, such as GPT-4, Llama and Mistral, has opened up new possibilities for human-machine interaction. This has led to the development of LLM-driven, text-based patient simulations in medical education. While these simulations allow students to practise communication skills, clinical practice involves richer interactions that go beyond text. For example, clinicians instruct patients to perform movements or observe their responses to actions, which are crucial for accurate diagnoses.

This paper introduces a novel framework to enhance OSCE training simulators, addressing these limitations in two key areas.

% PLOT: System in Action
\begin{figure*}[ht]
 \centering 
 \includegraphics[width=\linewidth]{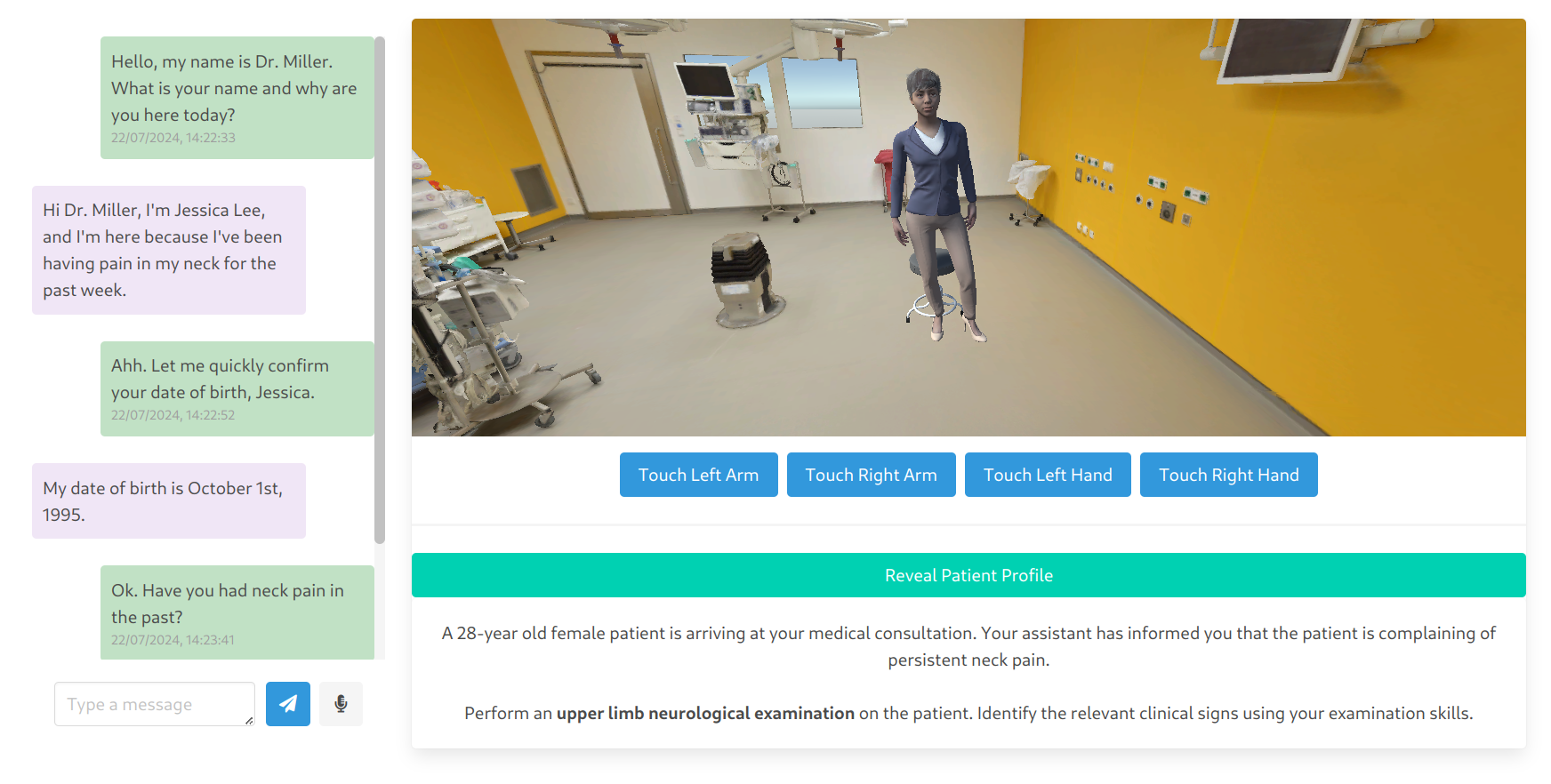}
 \caption{\textbf{The Basic Clinical Skills Training Simulator Interface.} This example illustrates a training session, showcasing: (Left) the multi-modal chat window for student interaction (voice or text) with the simulated patient; (Top-Right) the interactive Unity-based virtual patient capable of performing actions; and (Bottom-Right) the OSCE task description guiding the student.}
\label{fig:basic_clinical_skills_training_simulator_example}
\end{figure*}

\begin{itemize}
\item \textbf{Enhanced Patient Behavior Simulation.} We have extended the capabilities of patient simulators beyond text by introducing functional \textit{action spaces}. This enables simulated patients (whether virtual or robotic, as our framework is implementation-agnostic) to perform movements and actions, resulting in more realistic and interactive encounters. %Figure~\ref{fig:basic_clinical_skills_training_simulator_example} illustrates our simulator with a virtual patient.
\item \textbf{Advanced Automatic Feedback Generation.} We move beyond keyword-based scoring by using LLMs' understanding of language to provide comprehensive feedback. We introduce \textit{virtual tutors} that interact with students, respond to queries and provide detailed post-session evaluations. This dialogue-based assessment provides timely, targeted feedback to enhance student engagement and learning.
\end{itemize}

To validate our framework, we conducted a multifaceted evaluation. First, we assessed the objective performance of the system, including the critical real-time applicability metrics of latency and accuracy for the speech-to-text and language modelling components. Secondly, in a preliminary study, medical experts evaluated the quality of the patient simulation, focusing on the naturalness and coherence of the responses in a clinical skills scenario. Thirdly, these experts provided initial feedback on the quality of the automatically generated feedback, particularly in terms of its alignment with professional judgement. The aim of this work is to provide students with a more realistic, interactive and effective platform for preparing for the OSCE. In the long term, this framework serves as a blueprint for simulators that can deliver standardised, repeatable, and controlled training scenarios, thereby improving the consistency and reliability of OSCE evaluations.

\section{RELATED WORK}
\label{sec:related_work}
This section first provides an overview of the principles of OSCEs in medical assessment, before discussing technological advancements in medical education systems designed to improve OSCE training.

\subsection{Medical Background: Objective Structured Clinical Examinations}
\label{sec:background}
OSCEs provide an objective method for assessing the clinical skills of medical students through standardised, simulated clinical scenarios~\cite{harden1975assessment,khan2013objective}. Each scenario typically comprises the following:
a) a detailed \textit{patient profile} describing symptoms and behavior;
b) a \textit{task description} outlining the student's objectives; and
c) a \textit{checklist} of actions for objective evaluation by an examiner. OSCE scenarios, or stations, vary by clinical focus and student demands, but generally fall into four categories.

\begin{itemize}
\item \textbf{Clinical Examination:} Systematic investigation of a real or simulated patient.
\item \textbf{Clinical Procedures:} Performing skills like venipuncture.
\item \textbf{Communication Skills:} Interactions such as history taking or patient information delivery.
\item \textbf{Data Interpretation:} Analyzing X-rays or blood test results.
\end{itemize}

While in-person OSCE training with human-simulated patients (fellow students or professional actors) offers engaging experiences, it is often constrained by the high cost of actors and the limited variety of readily enactable scenarios. These limitations highlight the need for more scalable and versatile training solutions.

\subsection{Technological Advances in Medical Education}
\label{sec:medical_education_systems}
To address the challenges of traditional OSCEs, various technologies have been employed to simulate patients and enhance the learning experience, focusing on reducing costs and increasing accessibility.

\begin{figure*}[t]
 \centering 
 \includegraphics[width=\linewidth]{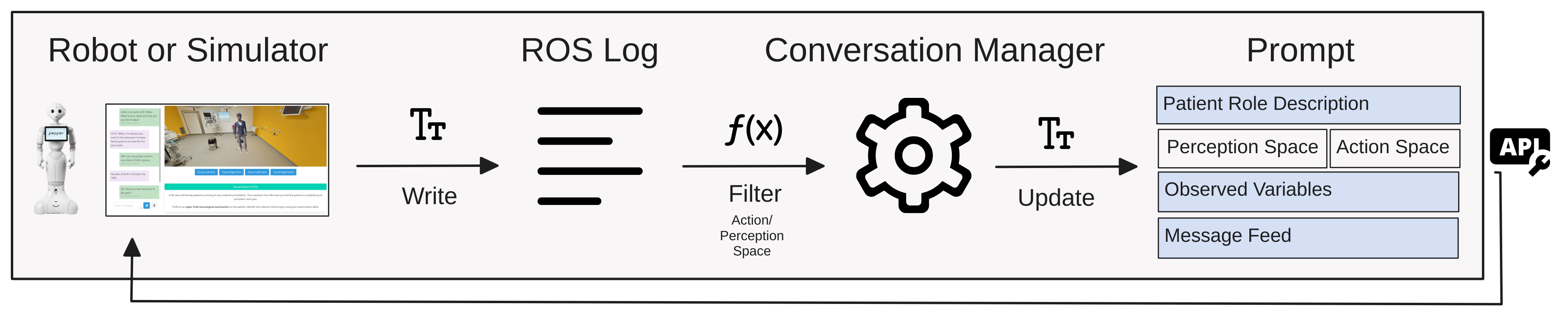}
\caption{\textbf{System Architecture Overview}: Data Flow for Dynamic Prompt Generation.
User interactions with the \textbf{Robot or Simulator} (frontend, e.g., physical robot or virtual environment) generate events (text, actions). These are written to a \textbf{ROS Log}-inspired message stream. The \textbf{Conversation Manager} (backend) filters this log based on the agent's defined \textbf{Action/Perception Space} functions. It then updates and constructs the dynamic \textbf{Prompt} for the LLM by incorporating the \textit{Patient Role Description}, \textit{Perception/Action Space Functions}, current \textit{Observed Variables}, and the recent \textit{Message Feed} (interaction history). The LLM's response then drives the subsequent behavior of the Robot or Simulator, completing the interaction loop.}
\label{fig:system_architecture_overview}
\end{figure*}

\textbf{Virtual Patient Simulation.}
Virtual patient simulators aim to replicate patient behavior for training in various clinical skills, including history taking~\cite{danforth2009development, CampillosLlanos2019DesigningAV} and clinical decision-making~\cite{McCallum2011ExploringNS}. Virtual reality environments can further enhance the perceived realism of these simulations~\cite{Rizzo2009AVR, Pensieri2014OverviewVR}. Historically, many virtual patients operated on hard-coded logic, limiting their response diversity to varied user inputs~\cite{CampillosLlanos2021LessonsLF}.

The advent of LLMs, with their advanced capabilities in nuanced role-playing and character consistency~\cite{zhu2025designing}, has significantly changed this landscape. LLMs are now integrated to create more dynamic text-based patient simulations, leveraging their extensive knowledge to portray diverse character traits and conversational abilities~\cite{Furlan2020ANL, park2023generative}. For instance, frameworks like CureFun~\cite{li2024leveraging} specifically leverage LLMs as simulated patients for clinical education, facilitating natural conversation practice and providing feedback on students' inquiry skills. Some ambitious projects~\cite{steenstra2025scaffolding} even construct entire simulated medical environments where LLM agents representing patients, doctors, and nurses interact, allowing doctor agents to learn and evolve their diagnostic and treatment capabilities through simulated experience.

While these systems demonstrate the power of LLMs for complex medical simulations and text-based dialogue training, many remain primarily confined to textual interactions. Our framework distinctively focuses on extending patient simulation beyond text to include functional action spaces critical for many OSCE examination scenarios, directly addressing the need for interactive, non-textual patient behaviors in training for medical students.

\textbf{Robotic Patient Simulation.}
For skills involving physical procedures (e.g. venipuncture, suturing and ultrasound scanning), robotic simulators offer valuable hands-on training ~\cite{MarcosPablos2022MoreTS, huang2017robot, hashimoto2021voice}. Dental robots, for example, simulate patient behavior in a cost-effective way for dental training ~\cite{takanobu2006dental}, and other systems teach therapeutic examination techniques ~\cite{ishikawa2015assessment}, suturing ~\cite{solis2008development} and surgical skills ~\cite{Howard2022ValueOR}. These systems excel at simulating specific physical responses. Our work builds upon this by combining the structured, state-based behavioral aspects often seen in robotics with the nuanced, context-aware predictive power of LLMs. This enables more coherent and diverse simulated behaviors that extend beyond text-based interactions.

%\textbf{Tutoring Systems.}
%Automated tutoring systems aim to facilitate learning by providing immediate feedback and performance assessments ~\cite{Feng2021ASR}. A notable example is Khan Academy's 'Khanmigo', an LLM-powered chatbot that provides students with hints and feedback ~\cite{hadi2023large}. Research in this area has explored the generation of step-by-step instructional strategies for LLM tutors ~\cite{Sonkar2023CLASSAD} and their application in subjects such as mathematics ~\cite{Macina2023MathDialAD}. With regard to OSCEs specifically, ~\cite{sugamiya2019construction} developed a state graph-based system for automatic assessment, where progression is tracked via speech recognition and keyword matching. While this approach is effective for structured procedures, it is less suited to open-ended dialogue scenarios or complex skill evaluations. Drawing inspiration from such automated scoring systems, our work seeks to enhance their capabilities by integrating more sophisticated language understanding from LLMs to provide comprehensive, nuanced feedback applicable to a broader range of clinical interactions.

\textbf{Tutoring Systems.}
Automated tutoring systems strive to facilitate learning by offering immediate feedback and performance assessment~\cite{Feng2021ASR}. A prominent example is Khan Academy's \textit{Khanmigo}, an LLM-powered chatbot providing students with hints and feedback~\cite{hadi2023large}. Research in this area has explored generating step-by-step instructional strategies for LLM tutors~\cite{Sonkar2023CLASSAD} and their application in domains like mathematics~\cite{Macina2023MathDialAD}. The application of LLMs extends to broader skill training, including social skills like communication and conflict resolution, where AI agents can act as practice partners and mentors, providing tailored feedback~\cite{yang2024social}.

Specific to OSCEs,~\cite{sugamiya2019construction} developed a state graph-based system for automatic assessment, where progression is tracked via speech recognition and keyword matching. While effective for structured procedures, this approach is less suited for open-ended dialogue scenarios or complex skill evaluations. Other LLM-based systems for medical education, such as CureFun~\cite{li2024leveraging}, also incorporate mechanisms to evaluate student dialogue by offering suggestions. Our work is inspired by such automated scoring and feedback approaches but seeks to enhance their capabilities by using LLMs to interpret the entirety of the student-patient interaction (including non-textual actions), thereby providing nuanced feedback applicable to a broader range of clinical interactions without relying on pre-defined keywords.

\section{SYSTEM ARCHITECTURE}
\label{sec:system_and_features}
Our framework for enhanced clinical skills training is realized through an application divided into a \textit{frontend} and a \textit{backend}. The backend (Section~\ref{sec:backend_component_combined}) manages the core logic, utilizing an LLM to simulate dynamic patient and tutor agents. The frontend (Section~\ref{sec:frontend_component_combined}) provides the user interface, including speech-to-text capabilities and a virtual patient simulator built with Unity. Figure~\ref{fig:system_architecture_overview} provides a visual overview. A key design principle is frontend independence, which allows the system to interface with either a virtual simulator or a physical robot, provided that a generic communication protocol is adhered to.

\subsection{Backend: LLM-Powered Agent Simulation}
\label{sec:backend_component_combined}
The backend's primary role is to simulate patient and tutor agent behaviors using prompts that instruct an LLM—in our implementation, Google's Gemini 1.5 Flash model, chosen for its balance of speed and quality in real-time interactions. The dialogue history management in the agent architecture is inspired by the Robot Operating Systems' (ROS) publish-subscribe mechanism. This ensures platform modularity. The agents are accessible via a REST API.\\

\begin{figure}[t]
  \centering
  \includegraphics[width=0.9999\columnwidth]{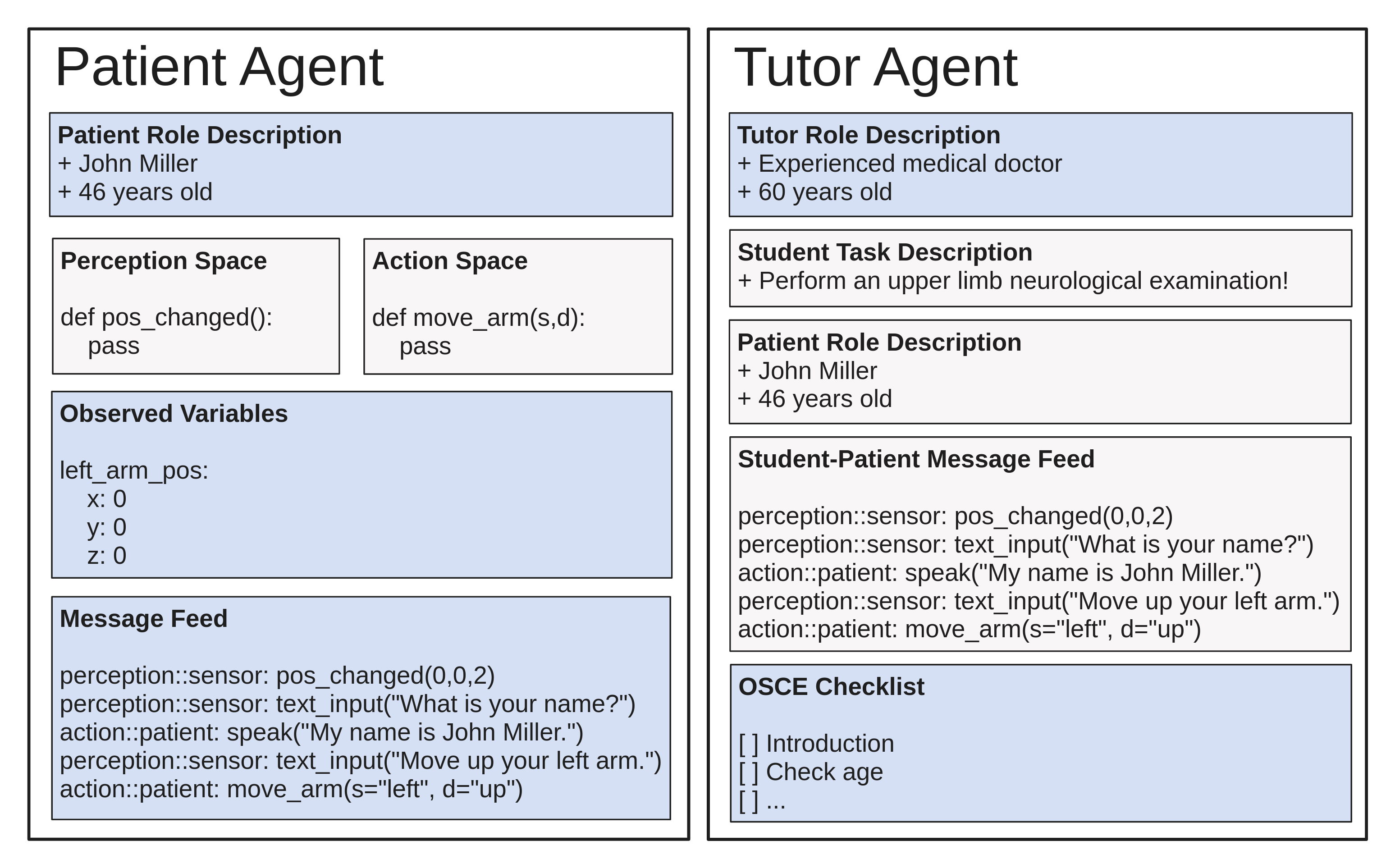}
    \caption{\textbf{Comparative Prompt Architectures for the Patient and Tutor Agents.}
    The figure details the distinct prompt structures fed to the LLM for (Left) the \textbf{Patient Agent} and (Right) the \textbf{Tutor Agent}. The Patient Agent's prompt includes its \textit{Role Description}, definable \textit{Perception} and \textit{Action Space Functions} (with example function signatures), current \textit{Observed Variables} (e.g., `\texttt{left\_arm\_pos}`), and a \textit{Message Feed} of recent interactions. The Tutor Agent's prompt incorporates its own \textit{Role Description}, the \textit{Student Task Description}, the \textit{Patient Role Description} (for context), the full \textit{Student-Patient Message Feed}, and the \textit{OSCE Checklist} used for evaluation.}
\label{fig:prompt_architectures}
\end{figure}

\textbf{Patient Agent: Enabling Dynamic and Interactive Behaviors.}
The patient agent simulates realistic patient responses during clinical scenarios. Its behavior, manifested as function calls executable by the frontend (virtual or robotic), encompasses motion, text, or speech. The agent's prompt (see Figure~\ref{fig:prompt_architectures}) is structured to facilitate natural and coherent interactions:

\begin{itemize}
    \item \textit{Patient Role Description:} Defines the patient's medical history, symptoms, and behavioral cues for the specific OSCE scenario.
    \item \textit{Perception Space:} %A list of functions triggered by sensor inputs (e.g., user text). These inputs, logged via the pub-sub mechanism, update the agent's context, allowing the LLM to generate relevant responses.
    A list of functions triggered by sensor inputs. These inputs, logged via the pub-sub mechanism, update the agent's context, allowing the LLM to generate relevant responses. For example, a perception space function such as \texttt{def on\_text\_input(text: str)} would log the student's utterance (e.g., \texttt{perception::sensor::text\_input("Please raise your arm.")}) for the LLM to process and incorporate into its situational awareness.
    \item \textit{Action Space:} %A list of functions the agent can execute (e.g., movements, speech). Logging these actions allows the LLM to consider recent behaviors when predicting subsequent actions. This directly enables Movement Behavior, where patients can be instructed to move limbs or adopt postures for examination, going beyond text-only interactions.
    A list of functions the agent can execute (e.g., movements, speech). Logging these actions allows the LLM to consider recent behaviors when predicting subsequent actions. For instance, an action space function like \texttt{def move\_arm(action\_type: str, parameters: dict)} would be logged for the LLM (e.g., \texttt{action::patient::move\_arm("left\_arm", \{"s":"left", "d":"up"\})}) and simultaneously trigger the corresponding animation in the Unity simulator. This directly enables \textbf{movement behavior}, where patients can be instructed to move limbs or adopt postures for examination, going beyond text-only interactions.
    \item \textit{Observed Variables:} A subset of relevant state variables (e.g., limb positions) made visible to the LLM, creating a notion of the agent's current state.
    \item \textit{Message Feed:} A log of recent  interactions of the pub-sub mechanism, providing a conversational and action history. The combination of \textit{Perception Space}, \textit{Observed Variables}, and the \textit{Message Feed} facilitates sophisticated reactive behavior. For instance, if a student manipulates the virtual patient's arm, the agent perceives this change via its observed variables and message feed, and can then respond appropriately (e.g., indicating pain based on its role description). Standard chat interaction for history taking and communication practice is also supported, building on established LLM capabilities.
\end{itemize}

\vspace{0.7em} % adjust (0.5em, 1em, etc.)

\textbf{Tutor Agent: Providing Comprehensive, Contextual Feedback.}
The tutor agent simulates an OSCE examiner, offering assistance and detailed evaluations. It can be invoked anytime and provides feedback via function calls executed on the frontend. Its core function is to deliver a textual evaluation and an OSCE score post-training, leveraging the LLM's understanding rather than simple keyword matching. The Tutor Agent's prompt includes:

\begin{itemize}
    \item \textit{Tutor Role Description:} Instructions on evaluation conduct, response style, and score calculation.
    \item \textit{Student Task Description:} The specific clinical task assigned to the student.
    \item \textit{Patient Role Description:} Provides the tutor with the same patient context the student has.
    \item \textit{Student-Patient Message Feed:} The complete transcript of the student-patient interaction, including all dialogue and actions.
    \item \textit{OSCE Checklist:} The critical list of steps the student should perform, with each correctly performed item typically scoring 1 point.
\end{itemize}

This design empowers several key tutor features:

\textbf{Interactive Help.} Students can query the tutor (e.g., "what next?"). The tutor, using its role description and knowledge of the student's task and patient profile, provides motivating, contextual hints. Scenario-specific knowledge can be added to the prompt for deeper support.

\textbf{Training Summary.} Post-session, the tutor generates a detailed summary highlighting positive and negative aspects of the student's performance. By analyzing the full \textit{Student-Patient Message Feed} against the \textit{OSCE Checklist}, the tutor identifies addressed, missed, or poorly executed steps. This offers far more nuanced feedback than keyword-based systems, assessing conversational quality and thoroughness.

\textbf{OSCE Score Calculation.} The tutor systematically reviews the \textit{OSCE Checklist} against the interaction log (inspired by chain-of-thought reasoning~\cite{wei2022chain}), assigning points for completed items. This LLM-based scoring offers significant advantages over keyword matching. LLMs can infer from conversational context whether checklist items are met, handling ambiguity better than keyword systems.  New OSCE scenarios can be created rapidly by providing only the role description, task description, and checklist, without needing to program new keyword logic. This is crucial for OSCE scenario management, where training scenarios stored in the backend can be easily updated.

\subsection{Frontend: User Interaction and Simulation Environment}
\label{sec:frontend_component_combined}
The frontend illustrated in Figure~\ref{fig:basic_clinical_skills_training_simulator_example} comprises:

\textbf{Chat Component.} Displays user, patient, and tutor messages. User messages prefixed with "@tutor" are routed to the tutor agent; tutor replies are color-coded.

\textbf{Speech Recognition.} In-browser voice interaction is enabled using a Whisper model~\cite{radford2023robust} via transformers.js~\cite{wolf2020transformers}, ensuring low latency.

\textbf{Virtual Patient Simulator.} Implemented in Unity and exported to WebGL. The rigged patient model executes behavioral functions (e.g. movements) defined in its action space and logs events using the ROS-inspired pub-sub mechanism.

\subsection{System Interaction Flow During a Training Session}
\label{sec:interaction_flow_combined}
A typical training interaction (see Figure~\ref{fig:system_architecture_overview}) begins with user input (text or voice) into the chat. Voice is transcribed by Whisper. The text is sent to the designated agent API (patient or tutor) in the backend. The backend's \textit{conversation manager} dynamically constructs the LLM prompt by:

\begin{itemize}
    \item Accessing the agent's defined action/perception spaces.
    \item Filtering the pub-sub log stream for relevant function calls and updating observed variables.
    \item Appending the recent interaction history (message feed).
    \item Including the agent's role description and action/perception function definitions.
    \item Inserting the current user query.
\end{itemize}

The LLM is tasked to complete the prompt, typically by generating a function call (with arguments) from the agent's action space in JSON format. This JSON is validated by the backend and sent to the frontend for execution by the virtual patient simulator or a connected robot. This loop continues throughout the OSCE training session.
%A comprehensive list of all prompts can be found in the Supplementary Material.

\begin{figure}[t]
  \centering
  \includegraphics[width=0.8\columnwidth]{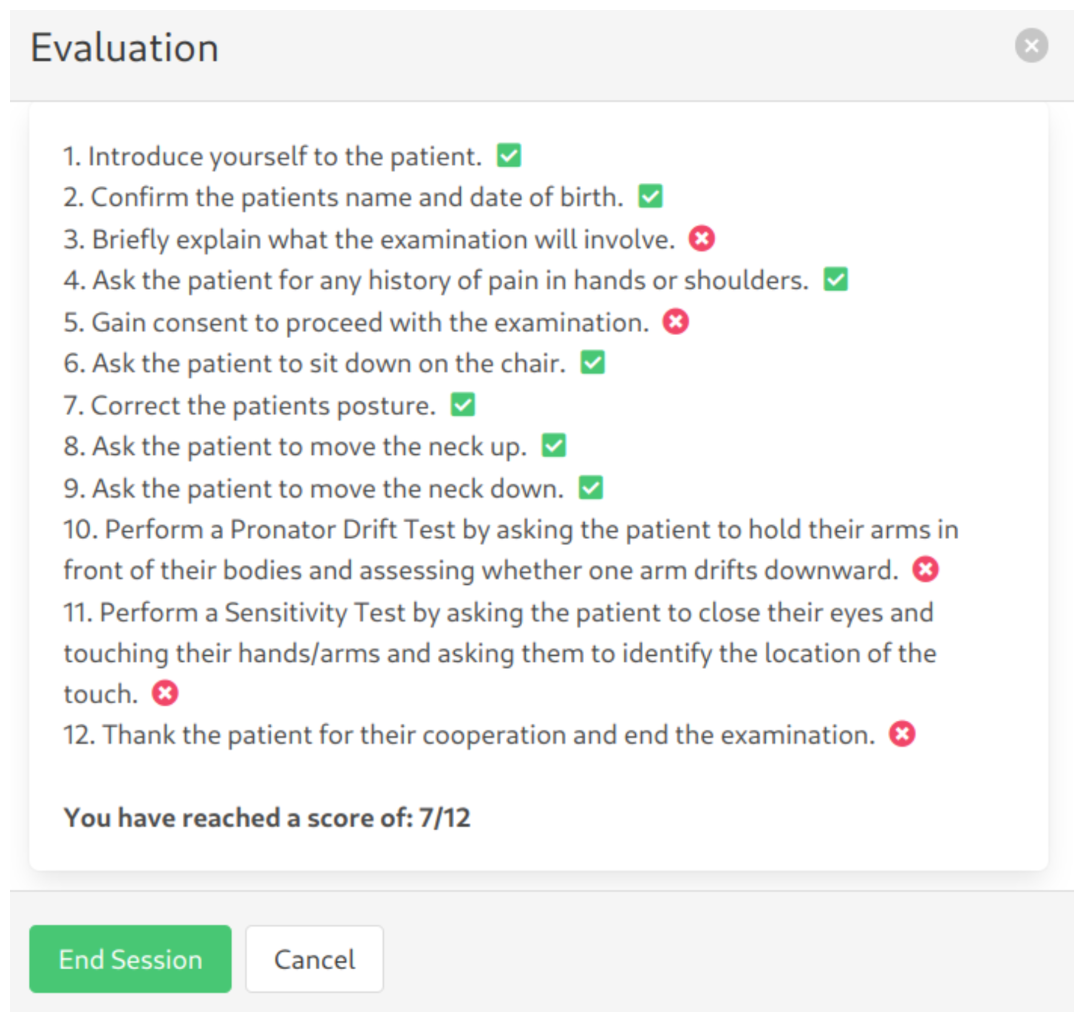}
\caption{\textbf{Example of the Virtual Tutor's Automated OSCE Score Evaluation.} The display shows a detailed checklist for an OSCE scenario, where each item is marked by the LLM-powered tutor as completed (green check) or missed (red cross) based on its analysis of the student-patient interaction.}
  \label{fig:tutor_osce_score}
\end{figure}
\section{EVALUATION}
\label{sec:evaluation}
This section presents an evaluation of our framework. We first assess quantitative performance benchmarks for critical system components (Section~\ref{sec:performance_benchmarks_combined}) and then discuss a preliminary expert evaluation focusing on the patient agent's behavioral fidelity and the tutor agent's feedback quality (Section~\ref{sec:expert_evaluation_combined}).

\subsection{Performance Benchmarks}
\label{sec:performance_benchmarks_combined}
Real-time interaction is crucial for an effective training system. We quantified the latency and accuracy of speech recognition and language model components to select optimal models for our application.

\subsubsection{Speech Recognition}
To select an optimal speech-to-text (STT) model, we compared three Whisper model versions~\cite{radford2023robust} on real-time factor (RTF; audio length to transcription time ratio) and word error rate (WER). Benchmarking used a LibriSpeech corpus subset~\cite{panayotov2015librispeech} (audio files $\leq$10s) on a standard Intel Core i7 CPU, reflecting our browser-based deployment.

Results are shown in Table~\ref{tab:table_performance_experiments}. The \textit{whisper-tiny} model (39M parameters) achieved the best RTF (0.325), indicating real-time transcription capability, with a WER (0.245) comparable to the larger \textit{whisper-base} (74M params, RTF 0.567, WER 0.231) and \textit{whisper-small} (244M params, RTF 1.91, WER 0.243) models. Given its significantly smaller size, faster loading, and superior real-time performance, \textit{whisper-tiny} was selected for our application. The \textit{whisper-small} model's RTF $>$1 disqualified it due to potential interaction delays.

\subsubsection{LLM Response Times}
We evaluated the response times of three LLMs: Gemini-1.5-Pro, Gemini-1.5-Flash (both Google Cloud-inferred~\cite{team2023gemini}), and the open-source, browser-executable Qwen-1.5-0.5B~\cite{bai2023qwen}. Each generated a 50-word story (approximating typical patient agent response length), repeated 10 times.

As detailed in Table~\ref{tab:table_response_time}, Gemini-1.5-Flash was the fastest (average response time: 2.07s ± 0.27), followed by Gemini-1.5-Pro (2.51s ± 0.13). The offline-capable Qwen-1.5-0.5B was substantially slower (14.82s ± 0.32). These results underscore that current offline models struggle to meet the $<$3-second response times generally desired for real-time interaction, unlike GPU-powered cloud models. Consequently, Gemini-1.5-Flash was chosen for its optimal balance of speed and quality. While cloud-based models have accessibility considerations, their performance is currently vital for this application, though future advancements may enable viable open-source, offline alternatives.

\begin{table}[t]
\caption{Speech Recognition Model Comparison}
\centering
\small % Reduce font size
\begin{tabular}{@{}lllllr@{}} % Adjust column widths and remove padding
\hline
Model & Real Time Factor & Word Error Rate & Params\\
\hline
whisper-tiny & $0.325$ & 0.245 & 39M \\
whisper-base & $0.567$ & 0.231 & 74M  \\
whisper-small & $1.91$ & 0.243 & 244M    \\
\hline
\end{tabular}
\label{tab:table_performance_experiments}
\end{table}

\begin{table}[t]
\caption{LLM Comparison}
\centering
\small % Reduce font size

\begin{tabular}{@{}lllllr@{}} % Adjust column widths and remove padding
\hline
Model & Response Time & Free Access & Offline \\
\hline
Qwen 1.5-0.5B-Chat & $14.82s \pm 0.32$ & \checkmark & \checkmark  \\
Gemini-1.5-Pro & $2.51s \pm 0.13$ & x & x  \\
Gemini-1.5-Flash & $2.07s \pm 0.27$ & x & x \\
\hline
\end{tabular}
\label{tab:table_response_time}
\end{table}

\subsection{Expert Evaluation: Patient Fidelity and Tutor Utility}
\label{sec:expert_evaluation_combined}
We conducted a preliminary evaluation with 18 medical professionals (8 experienced physicians ($>$5 yrs exp.), 4 medical assistants (3-5 yrs), 6 medical students ($<$3 yrs)) to assess the framework's practical utility.

%\textbf{Scenario and Procedure:} Participants engaged in a simulated OSCE scenario – an \textit{upper limb neurological examination} – lasting approximately ten minutes. This common task allowed for assessment of basic clinical skills.  
%Throughout the exam, experts rated the virtual patient's behavior (chat interaction, movement, reactivity) and the virtual tutor's feedback (chat help, summary, OSCE score appropriateness) on a 5-point Likert scale (1=poor, 5=very good) at each step.

%\textbf{Scenario and Procedure.} Participants engaged in a simulated OSCE scenario – an \textit{upper limb neurological examination} – lasting approximately ten minutes. This common task enabled the assessment of basic clinical skills.  To illustrate, checklist items included tasks such as introducing themselves and stating the purpose of the examination, inspecting the upper limbs for signs of wasting or abnormal posture, ensuring correct patient posture, and performing a pronator drift test (see Figure~\ref{fig:tutor_osce_score}). Throughout the examination, experts rated the behavior of the virtual patient (chat interaction, movement and reactivity) and the feedback of the virtual tutor (chat assistance, summary and appropriateness of OSCE score) on a 5-point Likert scale (1=poor, 5=very good) at each stage.

\textbf{Scenario and Procedure.} Participants engaged in a simulated OSCE scenario – an \textit{upper limb neurological examination} – lasting approximately ten minutes. This common task enabled the assessment of basic clinical skills. To illustrate, checklist items included tasks such as introducing themselves and stating the purpose of the examination, inspecting the upper limbs for signs of wasting or abnormal posture, ensuring correct patient posture, and performing a pronator drift test (see Figure~\ref{fig:tutor_osce_score}). Throughout the examination, experts rated the behavior of the virtual patient (chat interaction, movement and reactivity) and the feedback of the virtual tutor (chat assistance, summary and appropriateness of OSCE score) on a 5-point Likert scale (1=poor, 5=very good) at each stage. Alongside these quantitative ratings, qualitative feedback regarding specific interactions and suggestions for improvement was gathered from each expert through a structured post-session questionnaire and encouraged during brief, open discussions. This mixed-methods approach aimed to provide a comprehensive initial understanding of the system's perceived strengths and critical areas for future development.

\textbf{Overall Performance.} The virtual patient received a high average rating of 4.38 (SD=0.74), and the virtual tutor averaged 4.33 (SD=0.71). Experienced physicians rated the patient highest (4.56, SD=0.61), followed by students (4.43, SD=0.47) and assistants (4.01, SD=0.99). Tutor ratings were more consistent across groups (physicians: 4.38, SD=0.79; assistants: 4.38, SD=0.62; students: 4.16, SD=0.91). These scores suggest experts perceived the system as "good" to "very good."

\begin{figure}[t]
\centering
\includegraphics[width=\linewidth]{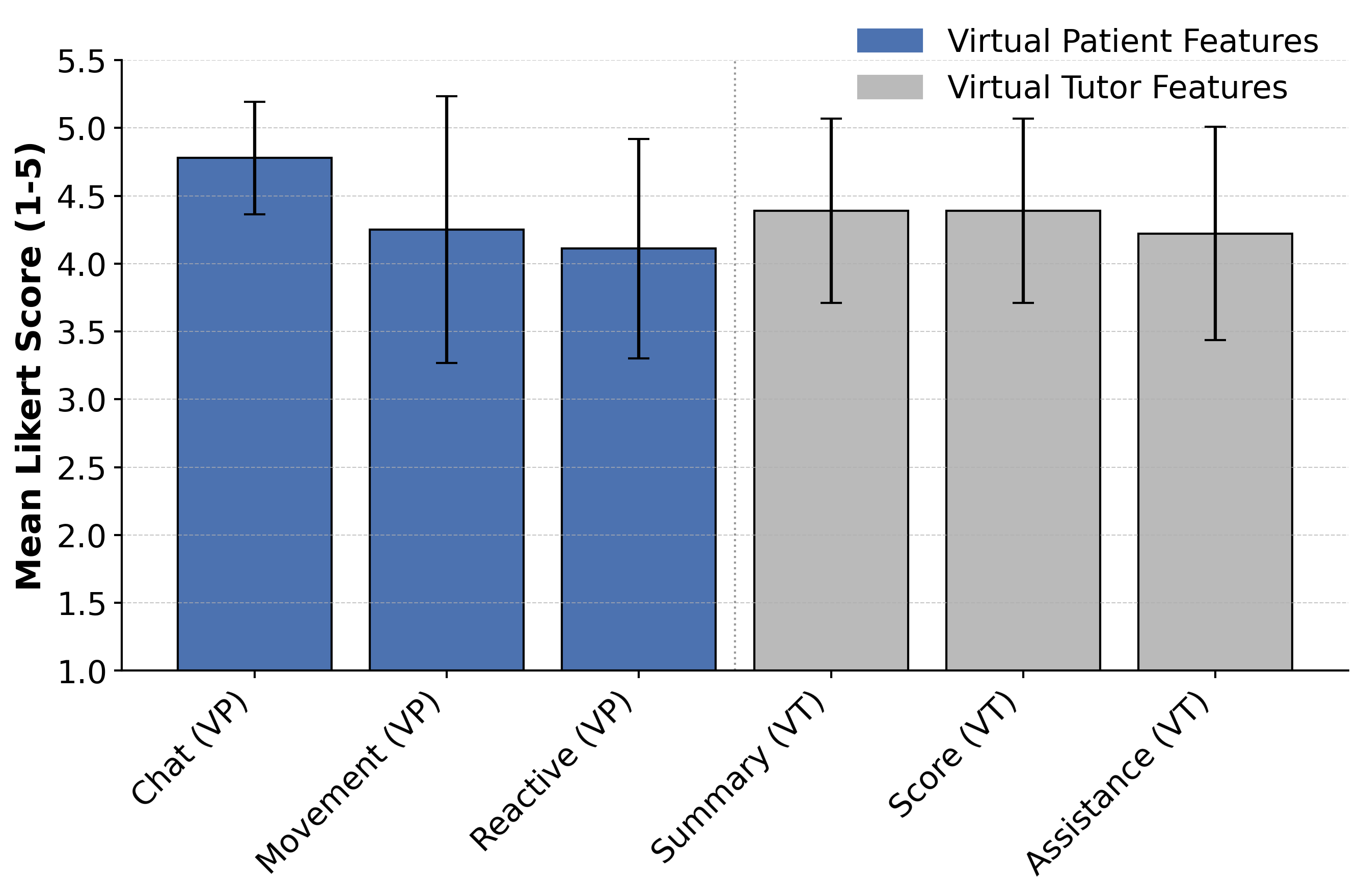} % Replace with your actual figure
\caption{\textbf{Mean Expert Ratings:} by Feature Category for Virtual Patient (VP) and Virtual Tutor (VT) performance. Ratings are on a 5-point Likert scale (1=poor, 5=very good). Error bars represent standard deviation. VP features include Chat, Movement, and Reactive behavior. VT features include Summary quality, OSCE Score appropriateness, and interactive assistance.}
\label{fig:category_ratings}
\end{figure}

\textbf{Categorical Performance and Qualitative Insights.} Figure~\ref{fig:category_ratings} details mean ratings per category. For the \textit{virtual patient}, \textit{chat interaction} was rated highest (M=4.78, SD=0.42). Experts praised its realism: "Wow, that was much more realistic than expected!" (E1) and responsiveness: "The patient reacts well to the commands!" (E7). \textit{Reactive behavior} (M=4.11, SD=0.81) and \textit{movement behavior} (M=4.25, SD=1.02) received slightly lower, though still positive, ratings. Feedback indicated areas for enhancement, such as simulating nuanced conditions ("Is it difficult to simulate the senile behavior of patients?" - E11) and handling complex instructions ("If possible, let the patient understand two commands in one sentence" - E5). For the \textit{virtual tutor}, the helpfulness of the \textit{summary} was rated highest (M=4.39, SD=0.68), while \textit{interactive assistance/help} during the scenario was rated slightly lower (M=4.22, SD=0.79). The experts showed a high agreement with the scores given by the virtual tutor (M=4.39, SD=0.68).
%The detailed summaries were appreciated for providing actionable feedback.

\textbf{Evaluation Summary and Future Directions.}
The results of the preliminary expert study illustrate our framework's potential for realistic OSCE training, particularly highlighting the success of LLM-driven dialogue generation for both patient interaction and tutor feedback summaries. The quantitative benchmarks confirm the feasibility of real-time performance with current cloud-based LLMs and optimized local STT.

Key areas for future development, informed by expert feedback, include enhancing the sophistication of the patient's reactive and movement behaviors, such as improving multi-command understanding and modeling more nuanced communication patterns associated with specific medical conditions. This involves refining prompts to elicit more subtle behavioral cues, thereby enriching the diagnostic challenge for students. Further work will also explore optimizing tutor interactions to provide even more effective in-scenario guidance. Addressing these points will further strengthen the realism and pedagogical value of the training system.

\section{CONCLUSION}
\label{sec:conclusion}
This paper introduced a novel framework leveraging LLMs to create dynamic patient and tutor agents, offering an enhanced digital training environment for OSCE preparation. Our primary contributions advance existing OSCE training methodologies by: (1) endowing simulated patients with movement and reactive behaviors that extend beyond text-only interactions, and (2) replacing simplistic keyword-based scoring with a comprehensive, LLM-driven evaluation of the entire student-patient interaction, culminating in a detailed summary and an interpretable OSCE score. Performance benchmarks confirmed the system's capability for real-time interaction, with core LLM responses averaging around 2 seconds. Crucially, preliminary evaluations by medical experts illustrated the pedagogical potential of our approach. They rated the virtual patient's behavior as notably natural and coherent, and the virtual tutor's feedback as helpful and appropriate. Future work will focus on several key areas: striving to integrate performant open-source, offline LLMs to improve accessibility and reduce operational costs; further refining patient simulations to capture the nuances of complex medical conditions and support more intricate action sequences; and continuing to enhance the sophistication of the virtual tutor's interactive guidance and assessment mechanisms. Ultimately, this line of research aims to provide medical students with increasingly realistic, accessible, and effective tools for mastering essential clinical skills.

%%%%%%%%%%%%%%%%%%%%%%%%%%%%%%%%%%%%%%%%%%%%%%%%%%%%%%%%%%%

%\clearpage
\bibliography{bibliography}
\bibliographystyle{IEEEtran}

%%%%%%%%%%%%%%%%%%%%%%%%%%%%%%%%%%%%%%%%%%%%%%%%%%%%%%%%%%%
% SUPPLEMENT
%%%%%%%%%%%%%%%%%%%%%%%%%%%%%%%%%%%%%%%%%%%%%%%%%%%%%%%%%%%

\clearpage
\includepdf[pages=-]{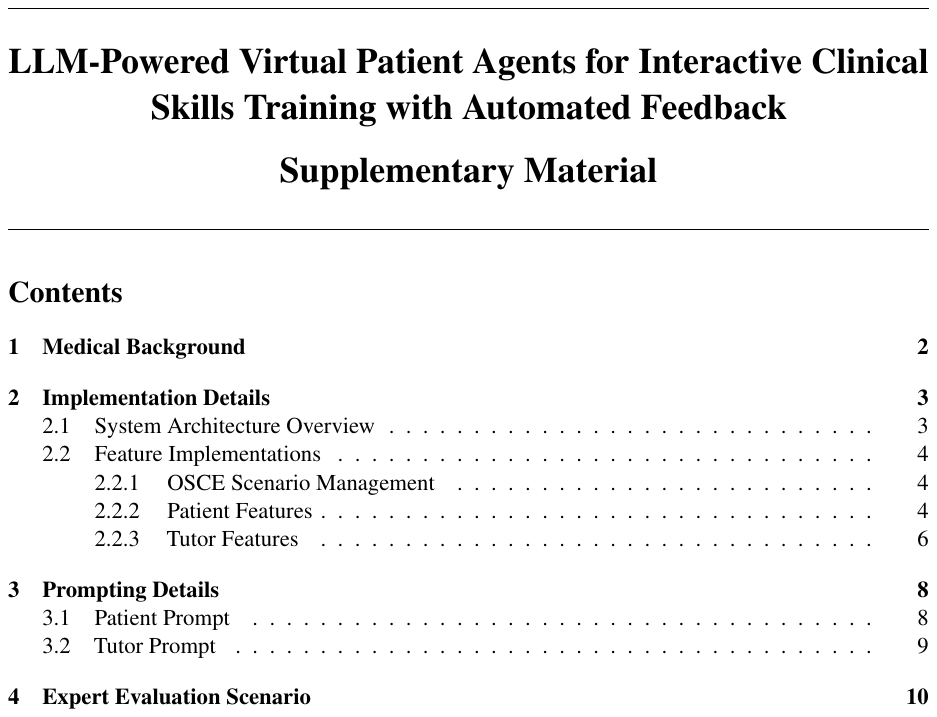}

\end{document}

\thispagestyle{empty}
\noindent
\rule{\textwidth}{0.3pt}
\begin{center}
  \textbf{\LARGE LLM-Powered Virtual Patient Agents for Interactive Clinical Skills Training with Automated Feedback\\\vspace{0.5cm}Supplementary Material}
\end{center}
%\textbf{\hfill submitted to NeurIPS 25}
\rule{\textwidth}{0.3pt}

\tableofcontents

\newpage

%%%%%%%%%%%%%%%%%%%%%%%%%%%%%%%%%%%%%
%%%%%%%%%%%%%%%%%%%%%%%%%%%%%%%%%%%%%
%%%%%%%%%%%%%%%%%%%%%%%%%%%%%%%%%%%%%
\section{Medical Background}
\label{sec:medical_background}

An Objective Structured Clinical Examination (OSCE) attempts to objectively assess the clinical skills of medical students using representative, simulated clinical \textbf{scenarios}~\citep{harden1975assessment,khan2013objective}. A scenario consists of the following artifacts: a) a detailed \textbf{patient profile}, which contains information about the patient's symptoms and behavior, b) a \textbf{task description}, which conveys what the student must do in this assessment, and c) a \textbf{checklist}, which contains a list of actions to be performed by the student. These actions are relevant to an objective assessment by an examiner, such as a medical professional.  OSCE scenarios, also known as stations, are conducted in a variety of ways depending on the clinical area they target and the demands they place on students~\citep{harden2015definitive}. However, most scenarios fall into one of the following four categories: 

\begin{itemize}
    \item \textbf{Clinical Examination} involves performing a systematic investigation on a real or simulated patient.
    \item \textbf{Clinical Procedures}, including the performance of a clinical skill such as venipuncture.
    \item \textbf{Communication Skills}, including interaction skills such as taking a patient's history or communicating information to a patient.
    \item \textbf{Data Interpretation}, such as the interpretation of X-rays or blood tests.
\end{itemize}

\noindent Nowadays, training courses and examinations based on the OSCE scheme are usually conducted in person. Patients are either simulated by fellow students or by professional actors. Although this provides an engaging learning experience for students, these forms of practice are very limited due to the costly use of actors and the small variety of scenarios that can be re-enacted.

\begin{figure}[thpb]
  \centering
  \includegraphics[width=0.8\columnwidth]{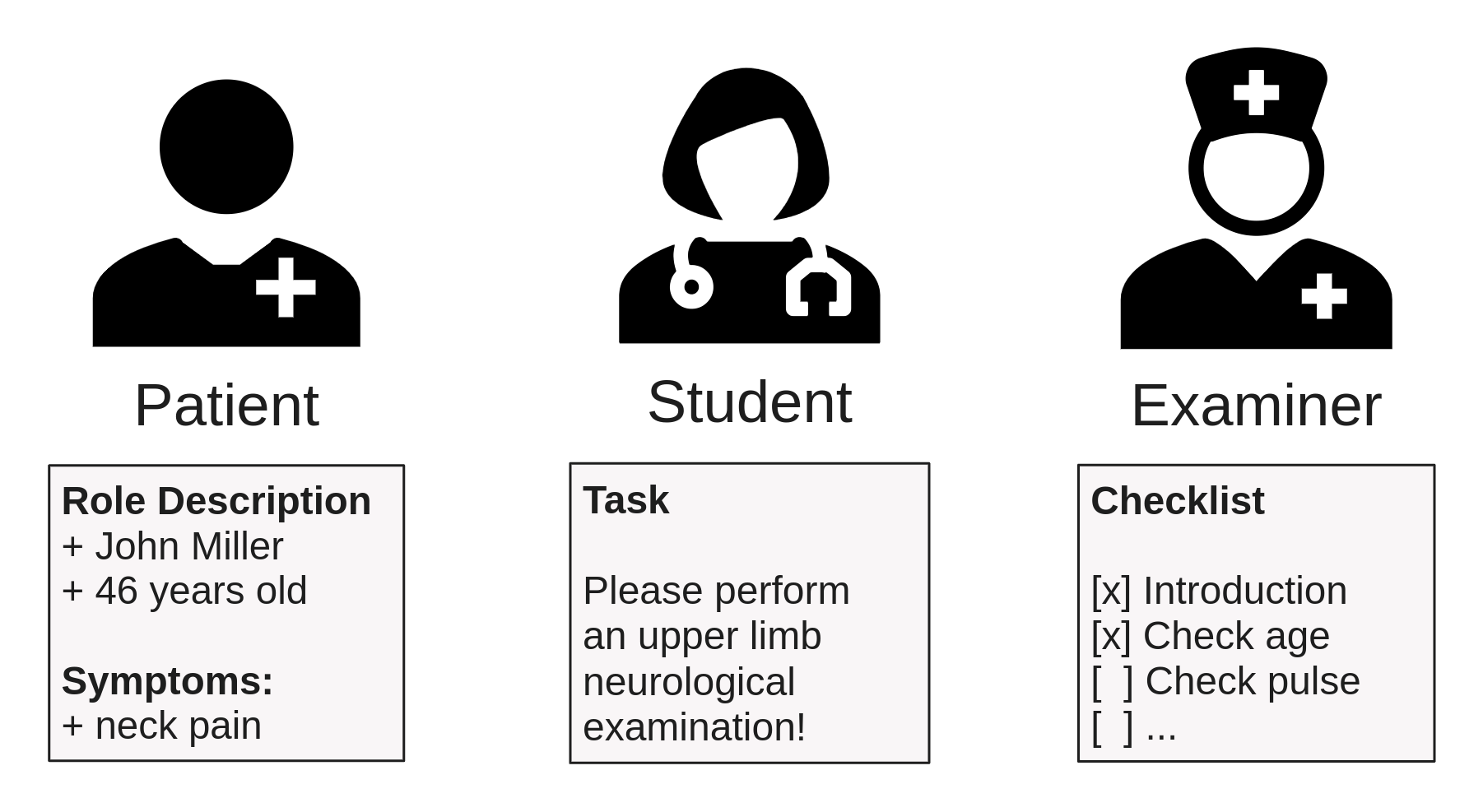}
  \caption{\textbf{Overview of an Objective Structured Clinical Examination (OSCE) scenario.} The figure illustrates the core components of a typical OSCE station. A \textbf{patient} is assigned a role with specific symptoms (e.g., John Miller, 46, with neck pain). A \textbf{student} is given a clinical task to perform, such as an upper limb neurological examination. An \textbf{examiner} uses a checklist to assess the student's performance on key actions.}
  \label{figurelabel}
\end{figure}

\section{Implementation Details}
\label{sec:implementation_details}

\subsection{System Architecture Overview}
\label{sec:system_architecture_overview}
\begin{figure}[thpb]
  \centering
  \includegraphics[width=0.8\columnwidth]{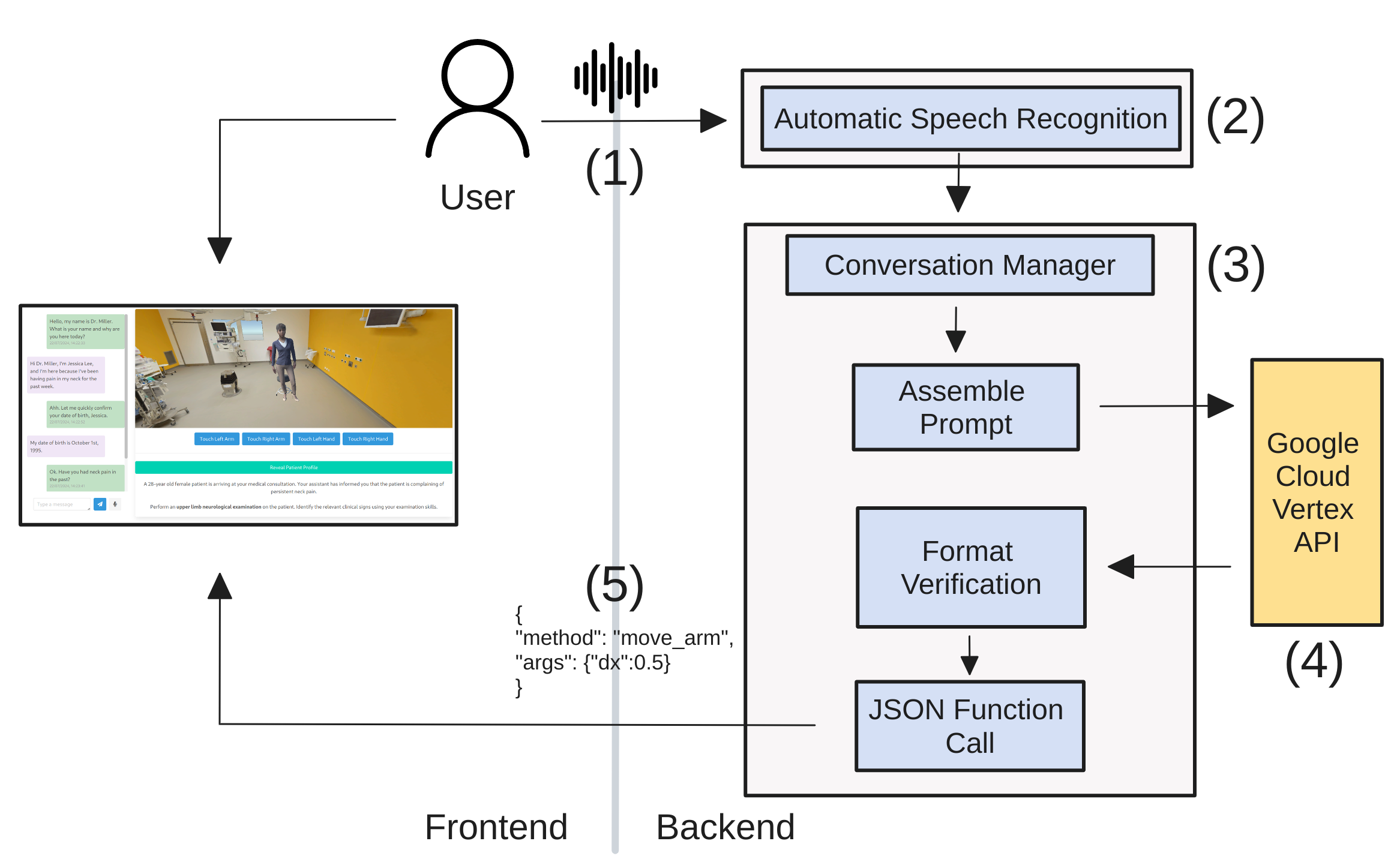}
  \caption{\textbf{System architecture overview.} The diagram illustrates the data flow for a user interaction. (1) A user interacts with the patient simulation via the frontend. (2) Spoken input is transcribed to text through Automatic Speech Recognition (ASR). (3) The conversation manager assembles the interaction history and current state into a prompt. (4) This prompt is sent to a Large Language Model (LLM) via the Google Cloud Vertex API, which generates a function call. (5) After format verification, the resulting JSON-RPC function call is executed by the frontend to update the simulation.}
  \label{fig:system_architecture_overview}
\end{figure}

To illustrate the functionality of our framework, we implemented a basic clinical skills training application. The system architecture is divided into two parts: the \textit{frontend} and the \textit{backend}. The \textit{backend} contains an API service that coordinates a language model to simulate the behavior of patient and tutor agents. The \textit{frontend} hosts a speech-to-text component that transcribes spoken words into text, as well as a virtual patient simulator implemented using the Unity game engine~\citep{haas2014history}. An overview of the system architecture is shown in Figure~\ref{fig:system_architecture_overview}. The backend of the system is not bound to a specific frontend and can be used either with a virtual patient simulator as the frontend or with a physical robot that then represents the patient, provided that the outlined generic communication protocol is used. 

% Dynamic Process Example 
\noindent The interaction of the simulator frontend with the backend during a training session proceeds as follows: The user either types or speaks an intended command into the chat component. If the microphone is used, the Whisper speech recognition model transcribes the audio into text. This text is then sent to the appropriate agent API (patient or tutor) in the backend component, depending on which agent is being addressed in the conversation. Once the message is received, the prompt is constructed. The dynamic prompt construction process is performed by the \textit{conversation manager}. The conversation manager has access to the function sets of the agent's action and perception space. It then filters the stream of logs resulting from the pub-sub mechanism from the simulator for these functions. When a change in the observed variables is detected, the conversation manager updates the observed variables object with those changes (e.g., a change in the position of the arms, legs, or head) and writes them to the prompt. It then adds the last $20$ messages from the filtered log to the prompt to provide the language model with the recent history of the interaction. It also copies the role description and action/perception space functions into the prompt. Finally, the user query is inserted into the prompt. The task given to the LLM is to complete the prompt after the \textit{action::} token with an appropriate function and arguments from the action space. This allows the language model to re-enact the given role and return a function call in JSON-RPC format. Once the function call string is generated, it is validated in the backend and then sent to the frontend where it is executed (in this case, the virtual agent in the Unity-based simulator environment). This completes the interaction loop and a new cycle is initiated until the end of the OSCE training session. 

\begin{figure}[t!]
  \centering
  \includegraphics[width=0.8\columnwidth]{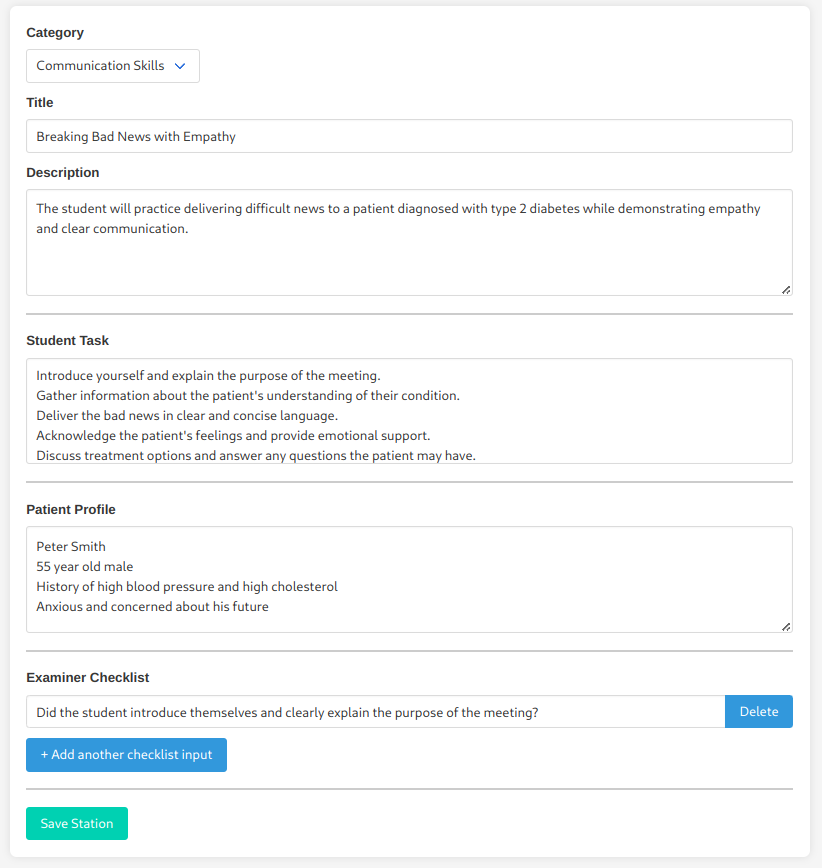}
  \caption{\textbf{Interface for creating custom OSCE scenarios.} Educators can author new training scenarios using this structured form. The figure shows the creation of a "Breaking Bad News with Empathy" scenario, which falls under the `Communication Skills` category. The author defines the steps of the \textbf{task} for the student, creates a detailed \textbf{role description} for the patient (including history and emotional state), and specifies the \textbf{OSCE checklist} items for the examiner to use during evaluation.}
  \label{fig:osce_scenario_management}
\end{figure}

\subsection{Feature Implementations}

\subsubsection{OSCE Scenario Management}
\label{sec:osce_scenario_managemtn}
The system allows the creation and management of OSCE scenarios, as illustrated in Figure~\ref{fig:osce_scenario_management}. To create a scenario for training, it is necessary to provide a patient's \textbf{role description} and a \textbf{task description} for the student. In addition, an \textbf{OSCE checklist} must be provided so that the tutor can perform an evaluation after the training. Training scenarios are stored in the backend and can be updated and modified at any time. In its standard configuration, the system covers the creation of customized OSCE scenarios for which the necessary perception and action functions of the underlying agent have been implemented.\\

\noindent The framework is designed to separate content creation from technical development to ensure scalability. The process is two-tiered:

\begin{itemize}
    \item \textbf{Educator:} A non-technical user can create a new scenario by providing text-based assets through a simple interface: the patient persona, the student tasks, the tutor persona, and the OSCE checklist. If materials are prepared, this can be done in a few hours.
    
    \item \textbf{Developer:} If a scenario requires a novel physical action not already in the library (e.g., a specific movement or gesture), a developer must code and animate this function. This is a more involved task, potentially taking a few days.
\end{itemize}

\noindent This modular design allows educators to rapidly create many new scenarios using the existing, extensive library of common physical movements, while providing a clear path for expansion.

\subsubsection{Patient Features}
\label{sec:patient_features}
The first goal of the proposed framework is to enhance the naturalness and coherence of the simulated patient character in training by extending its action capabilities beyond text to include interactive behavior. The following three features illustrate the capabilities of the simulated patient:\\ 

\noindent \textbf{Chat interaction.}
Chat interaction with the simulated patient allows for the practice of communication skills such as taking a patient's history or providing information to the patient. This has also been covered in previous work~\citep{danforth2009development, CampillosLlanos2019DesigningAV}. The student talks to the patient using either voice or text, and responses are returned in the chat.\\ 

\noindent \textbf{Movement behavior.}
By introducing an \textit{action space} of functions in our framework, simulated agents now support interactive movements in addition to pure text-based interaction. To perform an examination, patients can be instructed to move their arms, legs, or joints to certain positions and postures, and the students examine the trajectories.\\ 

\noindent \textbf{Reactive behavior.}
By introducing a \textit{perception space}, an internal state of \textit{observed variables} as well as a \textit{message feed} of the recent history of the interaction, the simulated patient is now capable of reactive behavior with respect to its internal state. For example, a student can change the position of the arm or push the patient's arm in a certain direction. The agent notices this change in position and, when asked, can tell if the movement is painful or not, depending on the medical condition specified.

\begin{figure}[t!]
  \centering
  \includegraphics[width=0.8\columnwidth]{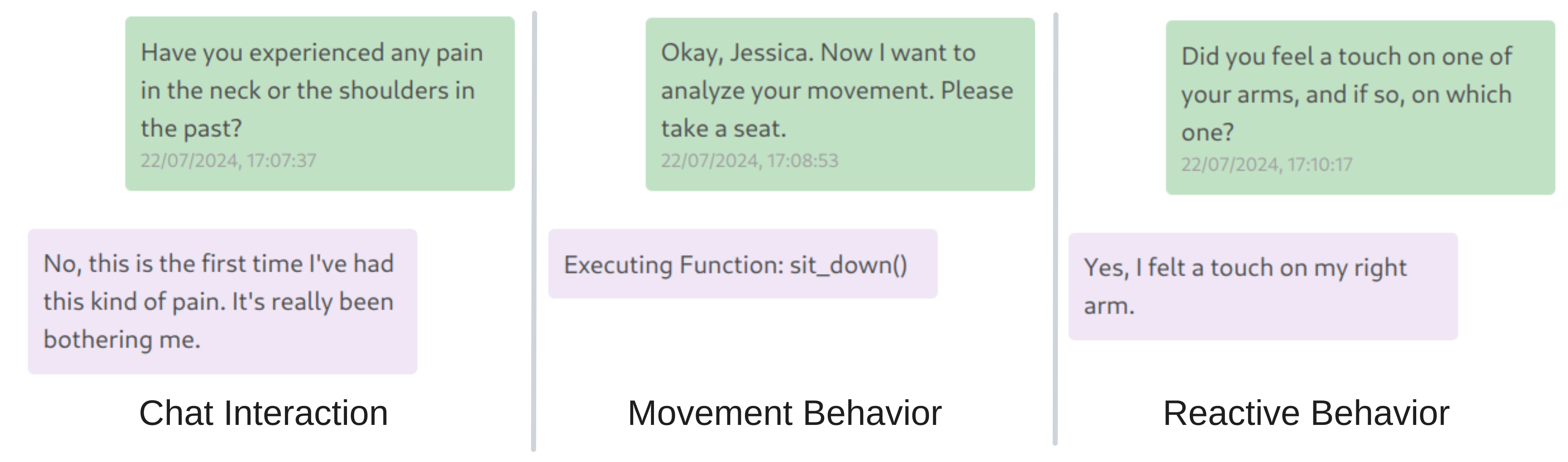}
  \caption{\textbf{Demonstration of the patient agent's multimodal behaviors.} This figure showcases the agent's ability to simulate a clinical encounter through three distinct interaction types. First, the agent conducts realistic \textbf{chat interactions} (left) to answer questions and in a conversation. Second, it translates commands into physical actions or \textbf{movement behaviors} (center), allowing the student to direct the patient during a training session. Third, the agent possesses \textbf{reactive behaviors} (right), where it reports on sensory input by accessing its internal state, providing crucial feedback for procedures like a neurological test.}

  \label{fig:patient_agent_features}
\end{figure}

\subsubsection{Tutor Features}
\label{sec:tutor_features}
The second goal of the proposed framework is to increase the detail and quality of feedback provided by identifying the steps performed in an OSCE checklist. This extends the keyword matching of existing work to a comprehensive performance assessment that encompasses the entirety of the interaction between the student and the patient agent. The following three features illustrate the capabilities of the tutor agent:\\ 

% Help throughout the Training
\noindent \textbf{Interactive help.}
At any point during a training session, students can ask the tutor for help, such as 'what step to take next'. The tutor is prompted with the patient's role description and knows the task the student is about to perform. In the tutor's role description, the tutor agent is encouraged to respond in a motivating way to help the student figure out how to proceed based on the information given. It is also possible to include additional scenario-related knowledge into the tutor agent prompt to allow for even more detailed case-based support and question-answering.\\ 

\begin{figure}[t!]
  \centering
  \includegraphics[width=0.99\columnwidth]{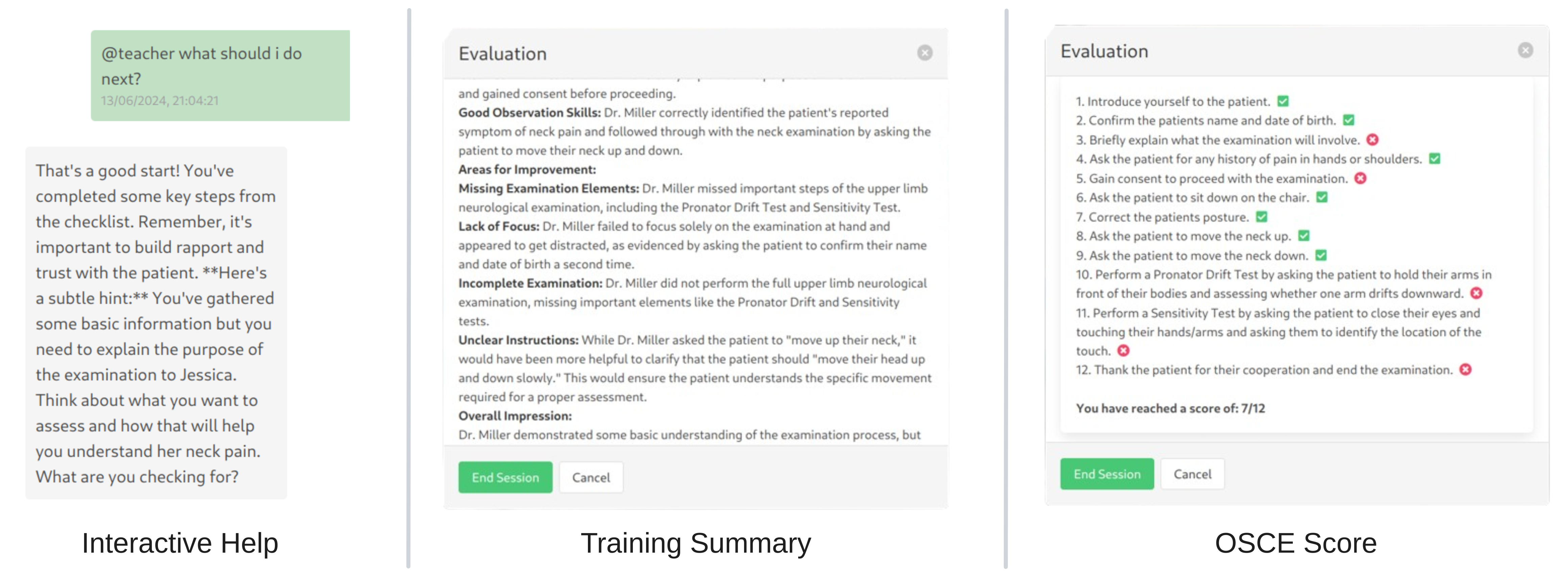}
  \caption{\textbf{Formative and summative feedback provided by the tutor agent.} The agent offers both on-demand guidance and post-session evaluation. (Left) During the simulation, it provides formative feedback through \textbf{interactive help}, guiding students with Socratic questions rather than direct answers. (Center and Right) After the session, it delivers summative feedback, including a \textbf{training summary} with qualitative observations and a detailed \textbf{OSCE score} that quantitatively measures performance against a checklist.}
  \label{fig:tutor_agent_features}
\end{figure}

% Summary 
\noindent \textbf{Training summary.}
After a training session is completed, the tutor agent provides a detailed summary of the training session. The summary is structured to list positive and negative aspects of the student's performance. Based on the OSCE checklist %the student's role description, and the task description, 
the tutor agent summarizes which checkpoints were appropriately targeted, which were forgotten, or which need to be improved in the future (see Figure~\ref{fig:tutor_agent_features}). By having access to the logs of the entire interaction, the agent can take into account everything that was said or done when summarizing the behavior. This allows for much more fine-grained feedback than a scoring approach that only checks whether certain keywords were used. In addition, by using the LLM's language understanding capabilities, the quality of the conversation can be evaluated much more deeply, allowing students to improve their conversational style and skills with patients.\\ 

% OSCE Score
\noindent \textbf{OSCE Score.}
Finally, after the training session has been summarized, the tutor agent distills the evaluation into an OSCE score. To do this, the agent is asked to think through the OSCE checklist step by step~\citep{wei2022chain}. Each successfully completed checklist item is worth 1 point, and each missed item is worth 0 points. The sum of all points is the OSCE score, which gives the student a concrete idea of how well it has performed. The student then knows which scenarios they are good at and which need further study. In contrast to a keyword-matching approach, calculating the OSCE score based on the entire interaction logs and the OSCE checklist has several advantages. First, new training scenarios can be generated very easily by providing only the artifacts \textbf{role description}, \textbf{task description} and \textbf{checklist}, and no hard-coded keyword matching has to be programmed. This makes the training application very \textbf{flexible} and \textbf{extensible}. Second, newly generated training scenarios can be \textbf{automatically evaluated} by LLMs because the language understanding capabilities of LLMs, in combination with the applied step-by-step evaluation procedure, allow them to infer from the context of the conversation whether an OSCE checklist item has been resolved or not. This is much more practical than approaches based on keyword-matching, since keywords may be ambiguous. 

%%%%%%%%%%%%%%%%%%%%%%%%%%%%%%%%%%%%%%%%%%%%%%%%

\clearpage  

\section{Prompting Details}
\label{sec:prompting_details}

\subsection{Patient Prompt}
The patient prompt used in the training scenario is as follows:\\

Imagine you're a real patient in an OSCE medical exam.

\vspace{1em}

\textbf{Your goal:} Clearly explain your symptoms and concerns to the doctor for an accurate diagnosis and treatment plan.

\vspace{1em}

\textbf{Background:}

\textit{\{role\_description\}}

\vspace{1em}

\textbf{Think like a real person:}
\begin{itemize}
    \item Describe your pain naturally, using your own words.
    \item Answer questions honestly and openly.
    \item React to the doctor's instructions and inquiries in a believable way based on your emotional state (use prompts if needed).
\end{itemize}

The more believable and natural your responses are, the better the learning experience will be for the medical student.

\vspace{1em}

\textbf{Remember:}
\begin{itemize}
    \item Use your best judgment to stay true to your character and situation.
    \item Feel free to ask clarifying questions if needed.
\end{itemize}

\vspace{1em}

\textbf{Patient State:}

\textit{\{observed\_variables\}}

\vspace{1em}

\textbf{Perception Space:}

\textit{\{perception\_space\}}

\vspace{1em}

\textbf{Action Space:}

\textit{\{action\_space\}}

\vspace{1em}

\hrule

\vspace{1em}

Please follow the instructions of the doctor and use only the actions from the Action Space provided to engage in the conversation effectively.

\vspace{1em}

\textbf{Current Conversation:}

\textit{\{history\}}

\newpage 

%%%%%%%%%%%%%%%%%%%%%%%%%%%%%%%%%%%%%%%%%%%%%%%%

\subsection{Tutor Prompt}
The tutor prompt used in the training scenario is as follows:\\

You are a seasoned medical teacher with over 20 years of experience, here to guide a student doctor preparing for their OSCE exam.

\vspace{1em}

\textbf{The Task at Hand:}

The student doctor is tasked with:

\textit{\{task\_description\}}

\vspace{1em}

\textbf{Your Role as a Mentor:}
\begin{itemize}
    \item You have access to the assessment checklist, which outlines key steps for the examination.
    \item The student doctor can request your help at any point.
    \item \textbf{Provide subtle hints} based on the checklist to nudge them in the right direction. Don't give away the answer!
    \item Focus on helping the student doctor develop essential skills like critical thinking, information recall, and effective communication.
    \item \textbf{Offer positive reinforcement} throughout the simulation to keep the doctor motivated.
\end{itemize}

\vspace{1em}

The assessment checklist for the examination is as follows:

\textit{\{checklist\}}

\vspace{1em}

\hrule

\vspace{1em}

\textbf{Current Conversation:}

The conversation between the student doctor and the patient is as follows:

\textit{\{history\}}

%%%%%%%%%%%%%%%%%%%%%%%%%%%%%%%%%%%%%%%%%%%%%%%%

\newpage

\section{Expert Evaluation Scenario}
\label{sec:evaluation_scenario}
\noindent This section contains a transcript of the OSCE scenario of the upper limb neurological examination used to conduct the expert evaluation. %The applied patient and tutor prompts and the OSCE checklist are illustrated in Section~\ref{sec:prompting_details}.
\newline

All results were obtained using the following 5-point Likert scale: 

\begin{itemize}
    \item (1) Strongly Disagree
    \item (2) Disagree
    \item (3) Neutral 
    \item (4) Agree
    \item (5) Strongly Agree
\end{itemize}

\vspace{1em}  % Adds 1 em of vertical space

%Participants were guided by the evaluation form to complete the following 5 tasks: 
The virtual exam contains the following 5 tasks: 
\newline

\noindent \textbf{Task 1: History Taking.}
Imagine you are a physician performing a brief neurological exam on a patient, including checking the function of the nerves in the neck and shoulders.

The first task in this training cycle is to take the patient's medical history.

An examination might include the following steps:

\begin{itemize}
    \item  Introducing yourself to the patient. 
    \item Confirming the patient's name and date of birth.
    \item Explain what the examination will entail. 
    \item Obtaining consent to proceed with the examination. 
\end{itemize}

\noindent \textit{Q1: The text answers given by the patient agent are reasonable and give the impression of a coherent character.}
\newline

\noindent \textbf{Task 2: Neck Posture Check.}
After taking the patient's history, our next step is to evaluate the posture and mobility of the neck.

An examination may include the following steps:

\begin{itemize}
    \item Having the patient sit down.
    \item Having the patient perform head movements and identify painful movements.
\end{itemize}

\noindent \textit{Q2: The patient agent executes the given commands as intended.}
\newline

\noindent \textbf{Task 3: Pronator Drift Test.}
Now we will execute a pronator drift test, which gives us insight about upper limb muscle weakness.

An examination may include the following steps

\begin{itemize}
    \item Having the patient stand up.
    \item Having the patient move their arms straight in front of their body while eyes closed.
    \item Identifying drifting directions in the arms and communicating the observations to the patient. 
\end{itemize}

\noindent \textit{Q3: The patient agent executes the given commands flawlessly.}
\newline

\noindent \textbf{Task 4: Sensitivity Test.}
In this part of the training we test if there are any differences in sensitivity between the two arms.

An examination may include the following steps

\begin{itemize}
    \item Having the patient stand up.
    \item Having the patient move their arms straight in front of their body while eyes closed.
    \item Touching the arms and/or hands of the patient and observing if the touched areas are sensitive. 
    \item Communicating the results to the patient. 
\end{itemize}

\noindent \textit{Q4: The patient executes the given commands well and gives reasonable answers to the requests.}
\newline

\noindent \textbf{Task 5: Training Summary.}
Now that you have completed the last step, it is time to get feedback from the Tutor Agent.

Please click the end session button on the right. 

A pop-up window will open and the Tutor will give you a summary of your performance. 

\noindent \textit{Q5: The summary given by the teacher is helpful and would help me as a student to improve.}
\newline

\noindent \textbf{Task 6: OSCE Score.}
The OSCE Score is an objective, transparent, checklist-based evaluation of a medical student's performance on a clinical skills examination.

To obtain the OSCE SCORE for the training you have completed, please click on the calculate OSCE score button.  

\noindent \textit{Q6: The OSCE score calculated by the teacher agent is fair and reasonable for my performance. }
\newline

\noindent \textbf{Task 7: Assistance.}
Please evaluate the quality of the help of the tutor during the training session. 

\noindent \textit{Q7: The teacher's answers to my questions were helpful and helped me to move forward during the training.}
\newline

%%%%%%%%%%%%%%%%%%%%%%%%%%%%
%%%%%%%%%%%%%%%%%%%%%%%%%%%%
%%%%%%%%%%%%%%%%%%%%%%%%%%%%

\clearpage
\bibliographystyle{plainnat}
\bibliography{bibliography}